\documentclass[10pt,conference,a4paper]{IEEEtran}
\usepackage{times,amsmath,epsfig}
\usepackage{epstopdf}

\usepackage{subfigure}
\usepackage{multirow}
\usepackage{graphicx}
\title{Spatial-Temporal Residue Network Based In-Loop Filter for Video Coding}
\author{%
{Chuanmin Jia{\small $^{\dag1}$}, {Shiqi Wang\small $^{*2}$}, {Xinfeng Zhang\small $^{\S3}$}, {Shanshe Wang\small $^{\dag4}$}, {Siwei Ma\small $^{\dag5}$}
 }%
\vspace{1.6mm}\\
\fontsize{10}{10}\selectfont\itshape
{\small $^{\dag}$}Institute of Digital Media, Peking University, Beijing, China\\
\fontsize{9}{9}\selectfont\ttfamily\upshape
$^{1,4,5}$\,\{cmjia,sswang,swma\}@pku.edu.cn
\vspace{1.2mm}\\
\fontsize{10}{10}\selectfont\rmfamily\itshape
$^{*}$\,Department of Computer Science, City University of Hong Kong, Hong Kong, China\\
\fontsize{9}{9}\selectfont\ttfamily\upshape
$^{2}$\,shiqwang@cityu.edu.hk
\vspace{1.2mm}\\
\fontsize{10}{10}\selectfont\rmfamily\itshape
$^{\S}$\,Rapid-Rich Object Search (ROSE) Lab, Nanyang Technological University, Singapore\\
\fontsize{9}{9}\selectfont\ttfamily\upshape
$^{3}$\,xfzhang@ntu.edu.sg
}
\begin{document}
\maketitle

\begin{figure}[b]
\parbox{\hsize}{\em

}\end{figure}

\begin{abstract}
Deep learning has demonstrated tremendous break through in the area of image/video processing. In this paper, a spatial-temporal residue network (STResNet) based in-loop filter is proposed to suppress visual artifacts such as blocking, ringing in video coding. Specifically, the spatial and temporal information is jointly exploited by taking both current block and co-located block in reference frame into consideration during the processing of in-loop filter. The architecture of STResNet only consists of four convolution layers which shows hospitality to memory and coding complexity. Moreover, to fully adapt the input content and improve the performance of the proposed in-loop filter, coding tree unit (CTU) level control flag is applied in the sense of rate-distortion optimization.
Extensive experimental results show that our scheme provides up to 5.1\% bit-rate reduction compared to the state-of-the-art video coding standard.
\\[1\baselineskip]
\end{abstract}


\section{Introduction}
%
Video compression is usually characterized by coding bits and perceived distortions of the reconstructed video.
To alleviate the distortions caused by video compression, many in-loop and out-loop filters have been studied for artifacts reduction of compressed videos \cite{norkin2012hevc} \cite{fu2012sample} \cite{zhang2017high} \cite{zhang2012adaptive}.
Norkin \textit{et al.} \cite{norkin2012hevc} proposed a deblocking algorithm to suppress discontinuities between adjacent blocks boundary.
Fu \textit{et al.} \cite{fu2012sample} developed a statistical method to train an offset at the encoder side and signal it to decoder to compensate the ringing effect induced by transform and quantization.
Zhang \textit{et al.} \cite{zhang2016low} \cite{ma2016nonlocal} incorporated the low rank algorithm into HEVC to develop non-local based adaptive loop filters.

Recently, convolutional neural network (CNN) achieved great success in image processing and video compression.
An artifacts reduction convolutional neural network (AR-CNN) algorithm was proposed for JPEG artifacts reduction by Dong \textit{et al.} \cite{dong2015compression}.
which boosted the restoration quality of JPEG coded images. Wang \textit{et al.} \cite{wang2016d3} also provided another network for the same task.
Park \textit{et al.} \cite{park2016cnn} directly integrated the super resolution convolutional neural network (SR-CNN) \cite{dong2014learning} into HEVC to replace deblocking filter and sample adaptive offset. However, this approach may lack its generalization ability because the same sequences were used for training and testing.
Dai \textit{et al.} \cite{dai2017convolutional} investigated the variable size transform of HEVC and proposed variable-filter-size residue-learning convolutional neural network (VR-CNN) for post processing in HEVC intra coding.
A decoder-end post-processing deep convolution network was investigated to boost the quality of decoded frames by Wang \textit{et al.} \cite{wang2017novel}.
Li \textit{et al.} \cite{li2017learning} designed an auto-encoder like CNN framework for image compression by using importance map for bit allocation.
Balle \textit{et al.} \cite{balle2016end} proposed an end-to-end framework form image compression and a novel method for rate estimation was proposed.

In this paper, we propose a spatial-temporal residue network (STResNet) based in-loop filter for HEVC inter coding by utilizing both spatial and temporal information.
In contrast with conventional CNN-based filter for video coding \cite{dai2017convolutional} \cite{park2016cnn},
the temporal information is also taken into account for in-loop filters to improve the quality of compressed frames.
In particular, the residue learning approach \cite{he2016deep} is adopted to accelerate the training process in our designed network and boost the coding performance.
To investigate the compatibility with the state-of-the-art video compression algorithm,
we integrate the proposed STResNet into HM 16.15 \cite{bossen2011hevc} as a novel inloop filtering method after sample adaptive offset (SAO).
Moreover, the coding tree unit (CTU) level control flags are designed to guarantee the performance of STResNet by rate-distortion optimization.
Experimental results indicate that STResNet provides higher visual quality and reduces on average 1.3\% bit-rate for HEVC in random access configuration.
The complexity and memory cost are also lower than the deep neural networks in \cite{dai2017convolutional}.

The organization of this paper is described as follows. Section 2 provides the details of the proposed STResNet in-loop filter.
Section 3 discusses the training process. Experimental results are shown in Section 4. Finally, Section 5 concludes this paper.

\section{Spatial Temporal Residue Network}
In this section, we mainly describe the detailed architecture of proposed spatial temporal residue network (STResNet) in-loop filters.
First, we discuss the network structure and parameters of STResNet. Then, the integration details of STResNet into HEVC are introduced.

\subsection{Network Structure}
\begin{figure}[h]
    \centerline{\psfig{figure=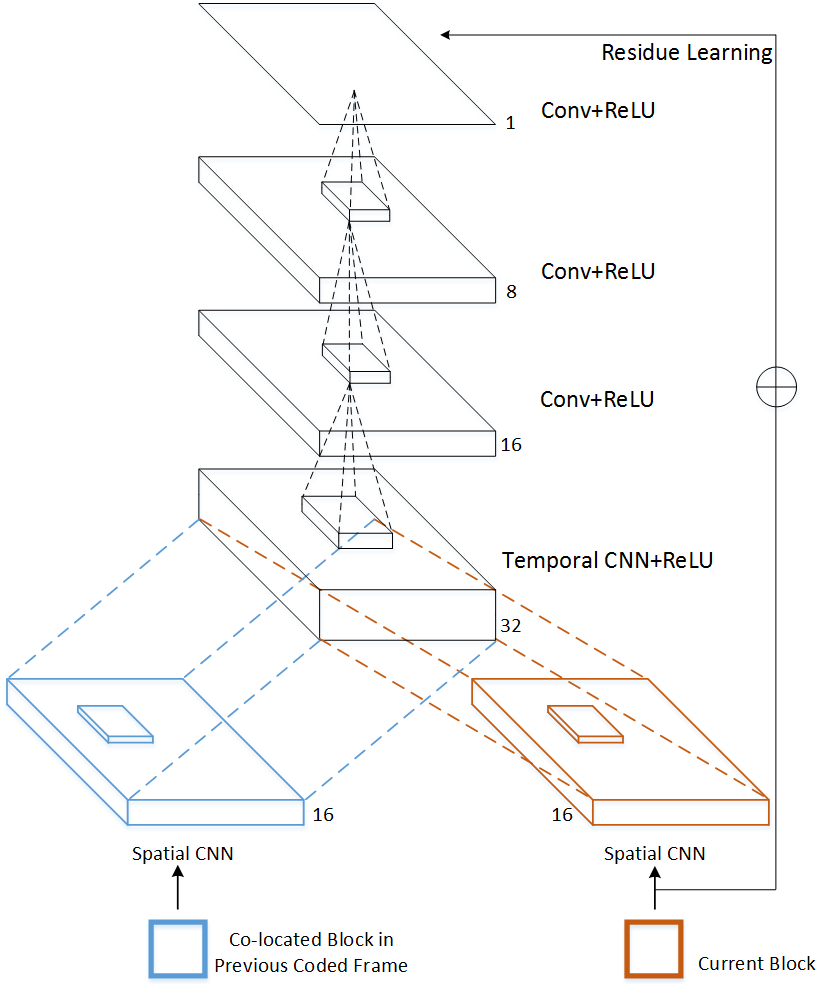,height=54.54mm,width=44mm} }
    \caption{The architecture of the proposed spatial temporal residue network.}\large
    \label{fig:netstruct}
\end{figure}

The structure of proposed STResNet is shown in Fig.~\ref{fig:netstruct}.
STResNet is a fully convolution network with four layers and also the feature map numbers of each layer are provided         in Fig.~\ref{fig:netstruct} and listed in Table~\ref{tab:param}.
There are two inputs for STResNet. One is the current block and the other is the co-located block of current block in the previous coded frame,
which is obtained from the closest reference frame of the current one.
The first layer contains two types of convolution operations which act as spatial CNN to extract spatial features respectively.
Filter sizes for spatial CNNs are $5\times5$ and $3\times3$.
For the temporal CNN layer, the outputs of the first layer are stacked for temporal CNN through fusion of feature maps and the filter size of which is set to be $3\times3$.
Two more convolution layers are followed by the temporal CNN layer, and both of them utilize $3\times3$ filters.
It is worth noting that rectified linear units (ReLU) \cite{nair2010rectified} are adopted as nonlinear mapping in STResNet for all convolution layers.
To accelerate the speed of training,
we design the four convolution layers as residue learning \cite{he2016deep} and the final output is the element-wise sum of current block and the output of the fourth convolution layer.
The detailed parameters of STResNet's are listed in Table~\ref{tab:param}.

\begin{table}[]
\centering
\caption{Descriptions of the parameters in STResNet}\footnotesize
\label{tab:param}
\begin{tabular}{|l|c|c|c|c|c|}
\hline
\multirow{2}{*}{}  & \multicolumn{2}{c|}{Layer1}                             & \multicolumn{1}{l|}{Layer2} & \multicolumn{1}{l|}{Layer3} & \multicolumn{1}{l|}{Layer4} \\ \cline{2-6}
                   & \multicolumn{1}{l|}{conv1} & \multicolumn{1}{l|}{conv2} & \multicolumn{1}{l|}{conv3}  & \multicolumn{1}{l|}{conv4}  & \multicolumn{1}{l|}{conv5}  \\ \hline
Filter Size        & $5\times5$                 & $3\times3$                 & $3\times3$                  & $3\times3$                  & $1\times1$                   \\ \hline
Feature Map Number & 32                         & 32                         & 16                          & 8                           & 1                           \\ \hline
Param Number       & 800                        & 288                        & 9216                        & 1152                        & 8                           \\ \hline
Total Param Number & \multicolumn{5}{c|}{11464}                                                                                                                        \\ \hline
\end{tabular}
\end{table}

\subsection{STResNet in HEVC}
We integrate the proposed STResNet into the latest version of HEVC reference software (HM 16.15) as an additional method of in-loop filter after sample adaptive offset (SAO).
To fully explore the performance of STResNet, we define the $64\times64$ coding tree unit (CTU) as our filtering unit, which means STResNet takes each CTU and co-located CTU in its closest reference frame as input.
To control on/off of the proposed in-loop filter, the rate-distortion (R-D) optimization strategy is employed by comparing the R-D cost $J_{1}$ and $J_{2}$
where
\begin{equation} \label{Eq:rdo}
J_{1} = D_{1} + \lambda R_{1}, J_{2} = D_{2} + \lambda R_{2}.
\end{equation}
Here, $D_{1}$ and $D_{2}$ indicate the distortion without the proposed in-loop filter and after the proposed in-loop filters respectively.
$R_{1}$ and $R_{2}$ denote the coding bits of the two scenarios. If $J_{1}>J_{2}$, then the flag is enabled and vice-versa.
Since in-loop filter does not introduce additional coding information, the rate term without proposed in-loop filters $R_{1}$ equals to the rate term after proposed filters $R_{2}$, such that only computing the distortion term (in terms of mean-square-error, MSE) is sufficient.

\section{Network Training}
This section discusses the training procedures of the proposed STResNet from three aspects.
First, the generation of training data is provided. Secondly, our training strategy is described.
Finally, our training hyper-parameters are listed.

\subsection{Generation of Training Data}
Each training sample is denoted as ($x_{i-1}$, $x_{i}$, $y_{i}$) where $x_{i-1}$ and $x_{i}$ represent the input current block and its co-located block in the closet reference frame, as illustrated in Fig.~\ref{fig:netstruct}. Moreover, $y_{i}$ denotes the original signal of $x_{i}$.
To generate the training data, we first compress three high-definition (HD) video sequences in AVS2 test sequences, (taishan$\_1920\times1080$, beach$\_1920\times1080$ and pkugirls$\_1920\times1080$) \cite{zheng2013avs2} by using HM 16.15 with random access (RA) configuration.
Since our STResNet is integrated after SAO, we turn ON both deblocking filter and SAO during compression of the three sequences. The first 100 frames of these three sequences are chosen to generate the training data.
Moreover, these three sequences are compressed using four different quantization parameters (QP: 22, 27, 32, 37) to generated training data for different model.
Note that we set all QP offsets to zero in RA configuration, which indicates all B frames are coded with the same QP as I frame.
Second, we set the block size of each training sample to be $38\times38$, which indicates that the size of $x_{i-1}$, $x_{i}$ and $y_{i}$ is $38\times38$.
Third, for the $i$-th frame ($i\in{2,3,4...,100}$), $x_{i}$ can be extracted in it and $x_{i-1}$ can be found at the same spatial index of the $i-$th frame's closest reference frame. $y_{i}$ denotes the original pixel of $x_{i}$. To obtain more training data, we extracted the training samples with 10 pixels overlap in the luminance channel of each frame.
Hence, we obtained 313,088 training samples for each QP.
Finally, we randomly shuffled all cropped training samples.

\subsection{Training Strategy}
To distinguish different quality levels, the networks are individually trained in terms of different coding configurations.
For training samples ($x_{i-1}$, $x_{i}$, $y_{i}$), the objective function of training is to minimize the following loss function Eqn.(~\ref{Eq:1})
where $\Theta$ represents the set of hyper parameters we use during training and $F(x_{i-1}, x_{i}|\Theta)$ denotes the STResNet.
\begin{equation} \label{Eq:1}
{\L}(\Theta)=\frac{1}{N}\sum_{i=1}^{N}\left \| F(x_{i-1}, x_{i}|\Theta)-y_{i} \right \|^{2}.
\end{equation}
We train our STResNet by using the deep learning framework Caffe \cite{jia2014caffe} on a NVIDIA TITAN X GPU.
Since there are four different QPs: 22, 27, 32, 37, one model is trained for each QP value.

For training configuration, zero padding is used for each convolution layer to ensure the same size of input and output images.
Gaussian random initialization with $0.001$ standard error is used to initialize all convolution weights and all bias terms are initialized with zero.
The batchsize of each iteration is set to be $128$. We use the first-order gradient-based optimization $Adam$ \cite{kingma2014adam} to train our objective functions Eqn.~\ref{Eq:1}.
The momentum of $Adam$ optimization are set to be 0.9 and the value of momentum2 is adaptive to different QP.
We use different learning rates when training models for different QPs.

\subsection{Hyper-parameters}
\begin{table}[]
\centering
\caption{Hyper-parameters for each QP}\footnotesize
\label{tab:hyper}
\begin{tabular}{|l|c|l|c|c|c|}
\hline
\multirow{2}{*}{}                                            & \multicolumn{2}{c|}{\multirow{2}{*}{QP=22}}                               & \multirow{2}{*}{QP=27}                               & \multirow{2}{*}{QP=32}                               & \multirow{2}{*}{QP=37}                                   \\
                                                             & \multicolumn{2}{c|}{}                                                    &                                                     &                                                     &                                                         \\ \hline
Base Learning Rate                                          & \multicolumn{2}{c|}{1e-6}                                                & 1e-7                                                & 1e-8                                                & 1e-8                                                    \\ \hline
\begin{tabular}[c]{@{}l@{}}Momentum\\ Momentum2\end{tabular} & \multicolumn{2}{c|}{\begin{tabular}[c]{@{}c@{}}0.9\\ 0.999\end{tabular}} & \begin{tabular}[c]{@{}c@{}}0.9\\ 0.990\end{tabular} & \begin{tabular}[c]{@{}c@{}}0.9\\ 0.988\end{tabular} & \begin{tabular}[c]{@{}c@{}}0.9\\ 0.988\end{tabular} \\ \hline
Iteration                                                    & \multicolumn{2}{c|}{600,000}                                             & 600,000                                             & 300,000                                             & 300,000                                                 \\ \hline
\end{tabular}
\end{table}
This subsection shows the hyper-parameters we utilize during our training process.
Table~\ref{tab:hyper} illustrates the basic learning rate, momentum of $Adam$ \cite{kingma2014adam} and total iteration time of each QP.
For QP 32 and 37, all the hyper-parameters are the same. With the deceasing of QP, we increase the base learning rate and other hyper-parameters, e.g., the momentum2.

\section{Experimental Results}
\label{sec:page style}
Extensive experimental results are provided in this section.
Specifically, the proposed STResNet is integrated into the HEVC reference software to act as a new method of in-loop filter after SAO.
In the proposed STResNet, only the luminance is processed using the trained model.
Moreover, CTU level control flags are used to maximize of performance of STResNet based on rate-distortion optimization.
In the RDO process, we only consider the distortion of luma channel.
Four typical quantization parameters are tested: 22, 27, 32, 37, and for each QP the corresponding network is used.
All experiments are conducted under random access (RA) configuration with zero QP offset for all B frames.
The perceptual visual quality comparison and brief complexity analysis are also discussed in the following subsections.
The first 100 frames of all test sequences are compressed for testing.

\subsection{Objective Quality}
To evaluate the coding efficiency, we compute the BD-rate \cite{bjontegaard2001calcuation} to show the performances of STResNet.
We test our coding performances of proposed STResNet on HEVC common test sequences \cite{bossen2011common} (from Class B to Class E) which are illustrated in Table~\ref{tab5:Allsequences}.
Experimental results of the proposed STResNet with CTU level control is shown in Table~\ref{tab5:Allsequences}. Anchor is HM16.15 with deblock and SAO turning ON.
STResNet achieves 1.3\% bit-rate savings on average for random access configuration. We report the coding performance of Y components since we only apply STResNet for luminance channel.
\begin{table}[t]
  \centering
  \renewcommand{\arraystretch}{1.2} \scriptsize
  \caption{Rate distortion performance of the proposed STResNet}\label{tab5:Allsequences}
  \begin{small}
  \begin{tabular}{|c|c|c|c|c|}
  \hline
  \multicolumn{2}{|c|}{\multirow{2}{*}{Sequences}}	&\multicolumn{1}{c|}{Random Access}	\\ \cline{3-3}
  \multicolumn{2}{|c|}{}                                               &Y	            	\\ \hline
  \multicolumn{1}{|c|}{\multirow{5}{*}{Class B}} &Kimono              & -0.7\%	     \\ \cline{2-3}
												&ParkScene            & -0.8\%	    \\ \cline{2-3}
												&Cactus               & -0.3\%	    \\ \cline{2-3}
												&BasketballDrive      & -1.0\%	    \\ \cline{2-3}
                                                &BQTerrace            & -0.1\%	    \\ \hline								
  \multicolumn{1}{|c|}{\multirow{4}{*}{Class C}} &BasketballDrill     & -1.0\%	   \\ \cline{2-3}
												&BQMall               & -1.1\%	   \\ \cline{2-3}
												&PartyScene           & -1.2\%	    \\ \cline{2-3}
												&RaceHorsesC          & -1.5\%	    \\ \hline
  \multicolumn{1}{|c|}{\multirow{4}{*}{Class D}} &BasketballPass      & -2.0\%	   \\ \cline{2-3}
												&BQSquare             & -1.8\%   \\ \cline{2-3}
												&BlowingBubbles       & -2.1\%	   \\ \cline{2-3}
												&RaceHorses           & -2.2\%	   \\ \hline	
  \multicolumn{1}{|c|}{\multirow{3}{*}{Class E}} &FourPeople          & -5.1\%	   \\ \cline{2-3}
												&Johnny               & -0.8\%	     \\ \cline{2-3}
												&KristenAndSara       & -1.5\%	    \\ \hline
  \multicolumn{2}{|c|}{\multirow{1}{*}{Overall}}                      & -1.3\%     \\ \hline
  \end{tabular}
  \end{small}
\end{table}

\subsection{Subjective Quality}
\begin{figure}[t]
\center
\subfigure[]{
\includegraphics[width=0.18\textwidth]{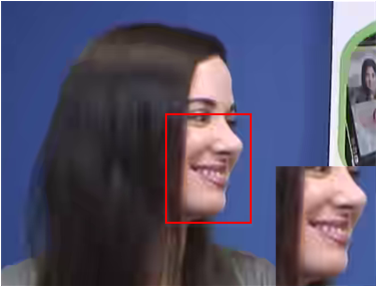}}
\subfigure[]{
\includegraphics[width=0.18\textwidth]{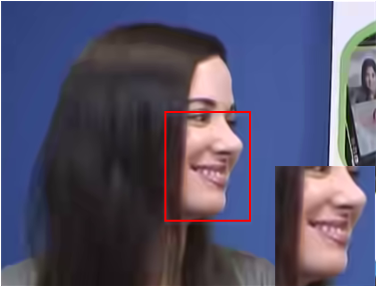}}
\footnotesize\caption{Visual quality comparison for $KristenAndSara$ at QP=37 (a) Anchor (b) Proposed.}
\label{Fig:subjective}
\end{figure}

In this subsection, we also compare the visual quality of reconstructed images as shown in Fig.~\ref{Fig:subjective}.
A crop of each image is enlarged in the bottom-right corner of each image.
It is obvious that the image processed by STResNet suppresses artifacts and distortions and produces better visual quality.

\subsection{Complexity Analysis}
To evaluate the time complexity of our algorithm, we test the proposed algorithm and record the encoding and decoding (enc/dec) time.
The information of testing environment is Intel i7 4770k CPU and CPU version Caffe \cite{jia2014caffe}.
Moveover, the operating system is Windows 10 64-bit home basic and the memory for the PC is 24 GB.
We illustrate the time complexity by using time increasing ratio.
\begin{equation} \label{Eq:time}
\Delta T = \frac{{T}'-T}{T}.
\end{equation}
where T is original enc/dec time and ${T}'$ is the proposed enc/dec time.
The proposed STResNet is integrated into HEVC reference software and the forward operation of network relies on libcaffe \cite{jia2014caffe}.
All of the time tests are conducted without GPU acceleration which means that we integrate the CPU version libcaffe \cite{jia2014caffe} into HM 16.15.
The ratio between encoding time of proposed STResNet in-loop filter and anchor encoding time is on average 135.7\%
while the ratio of decoding time is on average 703.1\% for all test sequences.
The details of enc/dec time for each class of HEVC common test sequences are listed in Table~\ref{tab:ratio}.
It is worth noting that the decoding time increase of Class D is greater than other class.
The explanation of this phenomenon is that the possibility of choosing STResNet among all CTU is higher than other classes.

\begin{table}[]
\centering
\caption{Time complexity of STResNet in-loop filter}
\label{tab:ratio}
\begin{tabular}{|l|c|c|}
\hline
           & $\Delta T_{enc}$  & $\Delta T_{dec}$ \\ \hline
Class B    & 137.5\%       & 481.1\%       \\ \hline
Class C    & 123.7\%       & 611.1\%      \\ \hline
Class D    & 127.9\%       & 1390.3\%       \\ \hline
Class E    & 153.7\%       & 895.6\%       \\ \hline
Average    & 135.7\%       & 703.1\%       \\ \hline
\end{tabular}
\end{table}

\section{Conclusion}

In this paper, a novel in-loop filter based on spatial and temporal residue learning (STResNet) is proposed for compensating the signal distortions and improving the visual quality of HEVC standard.
Specifically, the spatial-temporal coherences are jointly exploited to infer the pristine visual signal, such that the current frame can be reconstructed by feeding the current frame as well as the preceding ones into the spatial-temporal residue network.
In particular, the networks are individually trained in terms of different coding configurations to distinguish different quality levels and coding strategies.
To further explore the performances, the rate-distortion optimization strategy is employed in the proposed in-loop filter for CTU level control.
With the proposed STResNet in-loop filter, texture and high-frequency details can be efficiently restored, leading to better visual quality. Experimental results show that the proposed scheme achieves on average 1.3\% bit-rate reduction for all test sequences.

\section{Acknowledgement}
\footnotesize
This work was supported in part by the National Natural Science Foundation of China (61632001, 61421062), National Basic Research Program of China (973 Program, 2015CB351800), and the Top-Notch Young Talents Program of China, which are gratefully acknowledged.


%
%
%

\bibliographystyle{IEEEtran}


\end{document}